\documentclass{emulateapj}
\usepackage{graphics}

\newcommand{\spitzer}{\textit{Spitzer }}

\begin{document}


\shortauthors{TEMIM ET AL.}

\shorttitle{DUST PROPERTIES IN THE CRAB NEBULA} 

\title{PROPERTIES AND SPATIAL DISTRIBUTION OF DUST EMISSION IN THE CRAB NEBULA}

\author{TEA TEMIM\altaffilmark{1,2}, GEORGE SONNEBORN\altaffilmark{1}, ELI DWEK\altaffilmark{1}, RICHARD G. ARENDT\altaffilmark{1,3}, ROBERT D. GEHRZ\altaffilmark{4}, PATRICK SLANE\altaffilmark{5} AND THOMAS L. ROELLIG\altaffilmark{6}}

\altaffiltext{1}{Observational Cosmology Lab, Code 665, NASA Goddard Space Flight Center, Greenbelt, MD 20771, USA}
\altaffiltext{2}{Oak Ridge Associated Universities (ORAU), Oak Ridge, TN  37831, USA; tea.temim@nasa.gov}
\altaffiltext{3}{CRESST, University of Maryland-Baltimore County, Baltimore, MD 21250, USA}
\altaffiltext{4}{Minnesota Institute for Astrophysics, University of Minnesota, 116 Church Street SE, Minneapolis, MN 55455}
\altaffiltext{5}{Harvard-Smithsonian Center for Astrophysics, 60 Garden Street, Cambridge, MA 02138, USA}
\altaffiltext{6}{NASA Ames Research Center, MS 245-6, Moffett Field, CA 94035-1000, USA}

\slugcomment{Accepted by ApJ}

\begin{abstract}

Recent infrared (IR) observations of freshly-formed dust in supernova remnants (SNRs) have yielded significantly lower dust masses than predicted by theoretical models and measured from high redshift observations. The Crab Nebula's pulsar wind is thought to be sweeping up freshly-formed supernova (SN) dust along with the ejected gas. The evidence for this dust was found in the form of an IR excess in the integrated spectrum of the Crab and in extinction against the synchrotron nebula that revealed the presence of dust in the filament cores. We present the first spatially resolved emission spectra of dust in the Crab Nebula acquired with the Infrared Spectrograph aboard the \textit{Spitzer Space Telescope}. The IR spectra are dominated by synchrotron emission and show forbidden line emission from from S, Si, Ne, Ar, O, Fe, and Ni. We derived a synchrotron spectral map from the 3.6 and 4.5 $\micron$ images, and subtracted this contribution from our data to produce a map of the residual continuum emission from dust. The dust emission  appears to be concentrated along the ejecta filaments and is well described by an amorphous carbon or silicate grain compositions. We find a dust temperature of 55 $\pm$ 4 K for silicates and 60 $\pm$ 7 K for carbon grains. The total estimated dust mass is $(1.2-12)\times10^{-3}\:M_{\odot}$, well below the theoretical dust yield predicted for a core-collapse supernova. Our grain heating model implies that the dust grain radii are relatively small, unlike what is expected for dust grains formed in a Type IIP SN.

\end{abstract}


\section{INTRODUCTION} \label{intro}

Supernova (SN) dust is expected to form and has been observed to form shortly after the SN explosion in regions where the cooling gas is dense enough for grain growth to occur \citep[see review by][]{gal11}. The spectral signatures of this SN-formed dust depend on its temperature and composition, properties that are determined by the available gas abundances and the environmental conditions in the supernova remnant (SNR). Theoretical models based on classical nucleation theory predict that a range of dust species can form in SN ejecta, with some of the most abundant being MgSiO$_3$, SiO$_2$, Mg$_2$SiO$_4$, Si,
and C \citep[e.g.,][]{koz09}. The same models predict that the total dust mass produced per SN explosion should be 0.1-0.7 $\rm M_{\odot}$, with 0.04--0.2 $\rm M_{\odot}$ surviving the reverse shock and being recycled back into the interstellar medium (ISM) \citep[e.g.,][]{dwe08,koz09}. The large quantities of dust observed in high-redshift galaxies could be explained if an average SN produced 0.1--1.0 $\rm M_{\odot}$ \citep[e.g.,][]{dgj09}, but recent IR observations of dust in SNRs have yielded much lower masses.

Detailed studies of dust emission spectra from SN-formed dust have been limited to the closest  and youngest SNRs, mainly due to the lack of spatial resolution that is required to spatially separate ejecta dust from the swept-up ISM dust and confirm its SN origin. In recent years, spectroscopic observations of supernova remnants (SNRs) with the \textit{Spitzer Space Telescope} have confirmed the ejecta origin of SN dust in a few remnants by correlating its spatial distribution with ejecta gas \citep{rho09}. The observed emission spectra revealed spectral features that differ from those of typical ISM dust that predominantly consists of astronomical silicates and carbonaceous material, such as polycyclic aromatic hydrocarbons (PAHs), graphite, and amorphous carbon \citep[for a review see][]{zub04}. \citet{rho09} identified ejecta dust in the SNRs 1E102.2-7219, N132D, and Cas A, and found emission features from several dust species, including SiO$_2$, MgSiO$_3$, Si, SiC, Al$_2$O$_3$, Fe, and featureless carbon dust. IR emission spectra of ejecta dust in Cas A and SNR G54.1+0.3 revealed a broad 21 \micron\ dust feature that still remains unidentified \citep{rho08,tem10}.

The question of whether SNe are major contributors of dust to the ISM has still not been settled. The mass estimates of newly formed warm dust from recent \textit{Spitzer} observations are in the 0.02--0.1 $\rm M_{\odot}$ range \citep{sug06,rho08,rho09,tem10,bar10,sib10}. \textit{Herschel} observations of SN 1987A revealed 0.4-0.7 $\rm M_{\odot}$ of cool dust \citep{mat11}, while \textit{Spitzer} observations revealed up to 0.1 $\rm M_{\odot}$ of newly formed dust in SNR G54.1+0.3 \citep{tem10}. However, this dust will eventually encounter the SNR reverse shock. Without detailed information about the grain composition and mass, it is difficult to estimate how much of this dust will survive the encounter and be injected back to the ISM.

The Crab Nebula is a prototypical example of a pulsar wind nebula (PWN) that is sweeping up SN ejecta material, observed in the form of bright optical filaments \citep{gae06,hes08}. The evidence for the existence of dust in the Crab Nebula was found in the form of an IR excess above the PWN's synchrotron power-law spectrum \citep{tri77,gla82,mar84,dou01,gre04,tem06}. The IR excess observed in the integrated spectral energy distribution (SED) of the Crab implies a total dust mass of 0.001-0.03 $\rm M_{\odot}$, a range that depends on assumptions about dust composition and temperature. Optical observations  revealed absorption features from dust that spatially correlate with the cores of the optical filaments \citep{wol87,fes90, hes90, bla97,lol10}. Based on the measured dust-to-gas ratios in the filaments that range from 0.03--0.1 and the spatial correlation between dust extinction and low ionization lines, the dust in the Crab is most likely SN-formed dust that has been swept up by the PWN \citep{san98,lol10}. 

The properties of the dust in the Crab Nebula are not well determined. The IR continuum is dominated by synchrotron emission, and prior to \textit{Spitzer}, there were no detailed detections of dust emission spectra that could shed light on the grain composition. In this work, we removed the synchrotron contribution from spatially resolved \spitzer spectra of the Crab Nebula in order to isolate the dust emission spectrum for the first time and determine the physical properties and spatial distribution of the SN dust.

\section{OBSERVATIONS AND DATA REDUCTION} \label{obsv}

\begin{figure}
\epsscale{1.1} \plotone{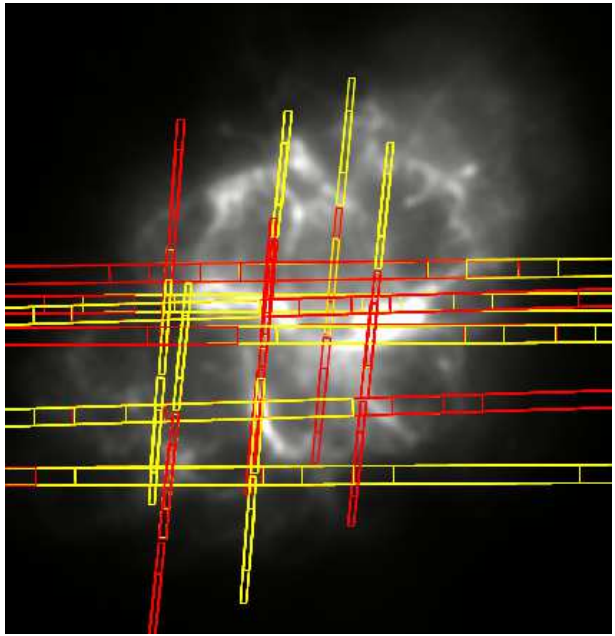}\caption{\label{slits}Positions of the IRS slits overlaid on the MIPS 24 $\micron$ image. The LL slits are the large slits oriented in the East/West direction and span the entire length of the SNR shell, and the SL slits are the narrower, North/South oriented slits.}
\end{figure}

The IR spectroscopy was carried out at six different positions across the Crab Nebula with all four modules of the \textit{Spitzer Space Telescope's} \citep{wer04,geh07} Infrared Spectrograph \citep[IRS,][]{hou04}; short-high (SH, 9.9--19.6 \micron), long-high (LH, 18.--9-37.2 \micron), short-low (SL, 5.2--14.3 \micron), and long-low (LL, 14.1--38.0 \micron). In this paper, we present sample SL and LL spectra. The observations were carried out between 2004 and 2006 under project ID 24, and processed with the pipeline version S18.7.0. They are summarized in Table \ref{obs}, including the coordinates, exposure times and AORKEYs. The corresponding slit positions are shown in Figure~\ref{slits}. A background spectrum was obtained with all four IRS modules at a position approximately 5\farcm4 southwest of the center of the nebula (see Table \ref{obs}). The spectra were extracted, calibrated and cleaned using the CUbe Builder for IRS Spectral Maps (CUBISM) version v1.7 \citep{smi07} and the background spectra for each module were subtracted from the source spectra. The background does not appear to vary spatially in the vicinity of the Crab Nebula, nor does it contain any strong emission lines that would contaminate the spectrum. Depending on the wavelength, the continuum level in the source spectrum is 3--7 times higher than the background level. The analysis was carried out with Spectroscopic Modeling, Analysis and Reduction Tool (SMART) version 8.1.2 \citep{hig04,leb11}.

The \textit{Spitzer} Infrared Array Camera (IRAC) observations at 3.6, 4.5, 5.8, and 8.0 \micron\ were carried out on 2004 March 6, while the Multiband Imagining Photometer for Spitzer (MIPS) observations at 24 and 70 \micron\ were carried out on 2004 March 14. The details about the reduction of the IRAC and MIPS data can be found in \citet{tem06}, where the images were previously published. All images in this paper are oriented such that North is up and East to the left.

\begin{deluxetable*}{lccccccc}

\tablecolumns{7} \tablewidth{0pc} \tablecaption{\label{obs}IRS OBSERVATIONS}
\tablehead{
\colhead{Position} & \colhead{RA/Dec (J2000)} & \colhead{AORKEY} & \colhead{Module} &  \colhead{Exposure/Cycle (s)}  & \colhead{Cycles}  & \colhead{Observation Date} 
}
\startdata
Position 1   &  5\fh34\fm31.00\fs    +22$^{\circ}$01$\arcmin$10$\farcs$0 &  3857664  &  SL (LL) &  480 (240) & 2 (4) & 2004-03-05   \\
	            

Position 2  & 5\fh34\fm29.55\fs    +22$^{\circ}$00$\arcmin$31$\farcs$0  &  16200704 & SL (LL)  & 60 (30) & 8 (8) & 2006-03-07  \\
	          
  
Position 3   & 5\fh34\fm33.88\fs    +22$^{\circ}$00$\arcmin$52$\farcs$0  &    16201216 &  SL (LL)  & 60 (30) & 8 (8) & 2006-03-07  \\
	            

Position 4  & 5\fh34\fm34.53\fs    +21$^{\circ}$59$\arcmin$46$\farcs$5  &    12634624 &  SL (LL)  &  60 (30) & 8 (8) & 2005-10-13 \\


Position 5   & 5\fh34\fm38.31\fs    +22$^{\circ}$00$\arcmin$49$\farcs$2  &   12634624  & SL (LL)  & 60 (30) & 8 (8) & 2005-10-13 \\


Position 6  & 5\fh34\fm38.44\fs    +21$^{\circ}$59$\arcmin$10$\farcs$0  &   16200704 &  SL (LL)  & 60 (30) & 8 (8) & 2006-03-07 \\


Background  & 5\fh34\fm39.91\fs    +22$^{\circ}$05$\arcmin$59$\farcs$6 & 16200704  & SL (LL) & 60 (30) & 8 (8) & 2006-03-07 \\
		     & 5\fh34\fm39.91\fs    +22$^{\circ}$05$\arcmin$59$\farcs$6 & 12634624 & SL (LL) & 60 (30) & 8 (8) & 2005-10-13 \\

\enddata
\tablecomments{The spectra at position 1 were previously published by \citet{tem06}.}
\end{deluxetable*}

\section{ANALYSIS AND RESULTS}\label{analysis}

\subsection{Subtraction of the Synchrotron Contribution} \label{synchanalysis}

The synchrotron emission from the Crab Nebula can be described by a power law spectrum of the form $L_{\nu}\propto\nu^{-\alpha}$, where $L_{\nu}$ is the synchrotron specific luminosity and $\alpha$ is the spectral index. At radio wavelengths, the spectral index is given by $\alpha=0.3$ \citep{baa77}, and steepens to $\alpha=0.7$ at optical wavelengths \citep{ver93}. This suggests that a break in the power-law spectrum occurs somewhere in the IR, between 10 and 1000 \micron\ \citep{mar84}. The most recent estimate for this break frequency is $(7.0^{+6.5}_{-3.4})\times10^2$ GHz, or $(4.3^{+5.7}_{-2.1})\times 10^2$ \micron, well beyond the \spitzer wavelength range \citep{are11}.

The first step in estimating the synchrotron contribution at each spatial position in the Crab Nebula was to create a map of the synchrotron spectral index $\alpha$,  We used the synchrotron-dominated IRAC 3.6 \micron\ and 4.5 \micron\ images to compute an index map, given by the equation
\begin{equation}
\alpha=\frac{ln(F_1/F_2)}{ln(\lambda_1/\lambda_2)},
\end{equation}
where $F$ is the surface brightness and $\lambda$ is the wavelength. The IRAC images were first background subtracted and corrected for extinction based on the $A_{\lambda}/A_K$ values from \citet{ind05}, and the relation $N_{H}/A_{K}=1.821\times10^{22}{\rm\ cm^{-2}}$ from \citet{dra89}. The hydrogen column density for the Crab was set to $N_H=3.54\times10^{21}\: \rm cm^{-2}$ \citep{wil01}. This led to correction factors of 1.11 and 1.08 for the 3.6 and 4.5 \micron\ channel, respectively. The images were also corrected by the IRAC surface brightness correction factors for extended sources; 0.91 for 3.6 \micron\ and 0.94 for 4.5 \micron. The 3.6 \micron\ image was then convolved to the resolution of the 4.5 \micron\ image using the convolution kernels developed by \citet{gor08}. The nominal wavelengths for the two channels that were used in the final calculation for the index are 3.550 \micron\ and 4.493 \micron, as specified by the IRAC Instrument Handbook.

\begin{figure}
\epsscale{1.2} \plotone{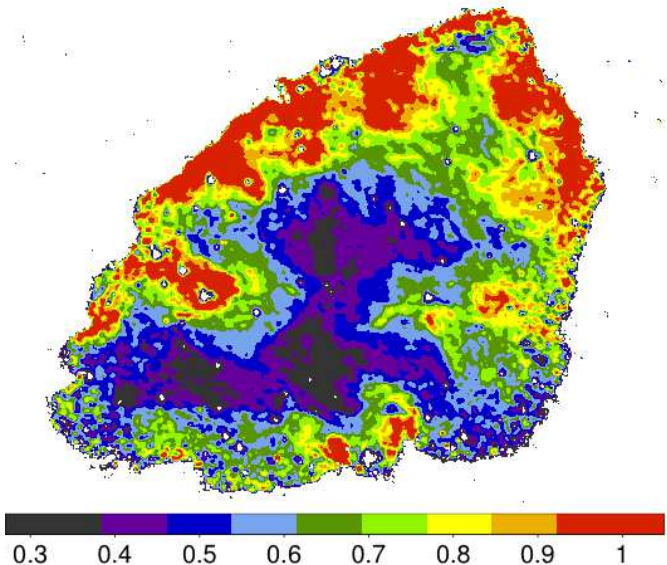} \caption{\label{indexmap} Synchrotron index map calculated from the synchrotron-dominated IRAC 3.6 and 4.5 \micron\ images. The spectral index $\alpha$ steepens from 0.3 in the torus and jet structures to 0.7 (1.0) in the outer SE (NW) parts of the nebula.}
\end{figure}

The final synchrotron index map is shown in Figure~\ref{indexmap}. Due to the updated extinction correction, the values of the index are slightly higher than those found in \citet{tem06}. The value of the synchrotron index in the inner torus and jet is approximately $\alpha=0.3$ and it gradually steepens towards the outer edges of the nebula due to synchrotron losses. The index steepens to about $\alpha=0.6$ towards the SE, and $\alpha=1.0$ towards the NW edge of the nebula. A similar asymmetry is seen in the spectral index map at optical wavelengths, where the index steepens to $\alpha\sim0.8$ in the SE, and exceeds $\alpha=1.0$ in the NW \citep{ver93}. The calibration uncertainty on the ratio of the fluxes in the two IRAC channels is approximately 4\% \citep{rea05}, which translates into a $\pm0.18$ uncertainty in the synchrotron spectral index. This large uncertainty in the spectral index is systematic across the nebula and does not affect the spatial variations in the in index.

In order to investigate the morphology of the dust and line emission in the Crab Nebula, we subtracted the synchrotron contribution from the IRAC 5.8 and 8.0 \micron\ images and the MIPS 24 and 70 \micron\ images. We produced a synchrotron-subtracted image for each wavelength by subtracting the synchrotron contribution extrapolated from the IRAC 3.6 \micron\ image using the equation
\begin{equation}
F_{i,residual} =F_i - F_{3.6} (3.550/\lambda_2)^{-\alpha}. \label{eq2}
\end{equation}

\begin{figure*}
\epsscale{1.0} \plotone{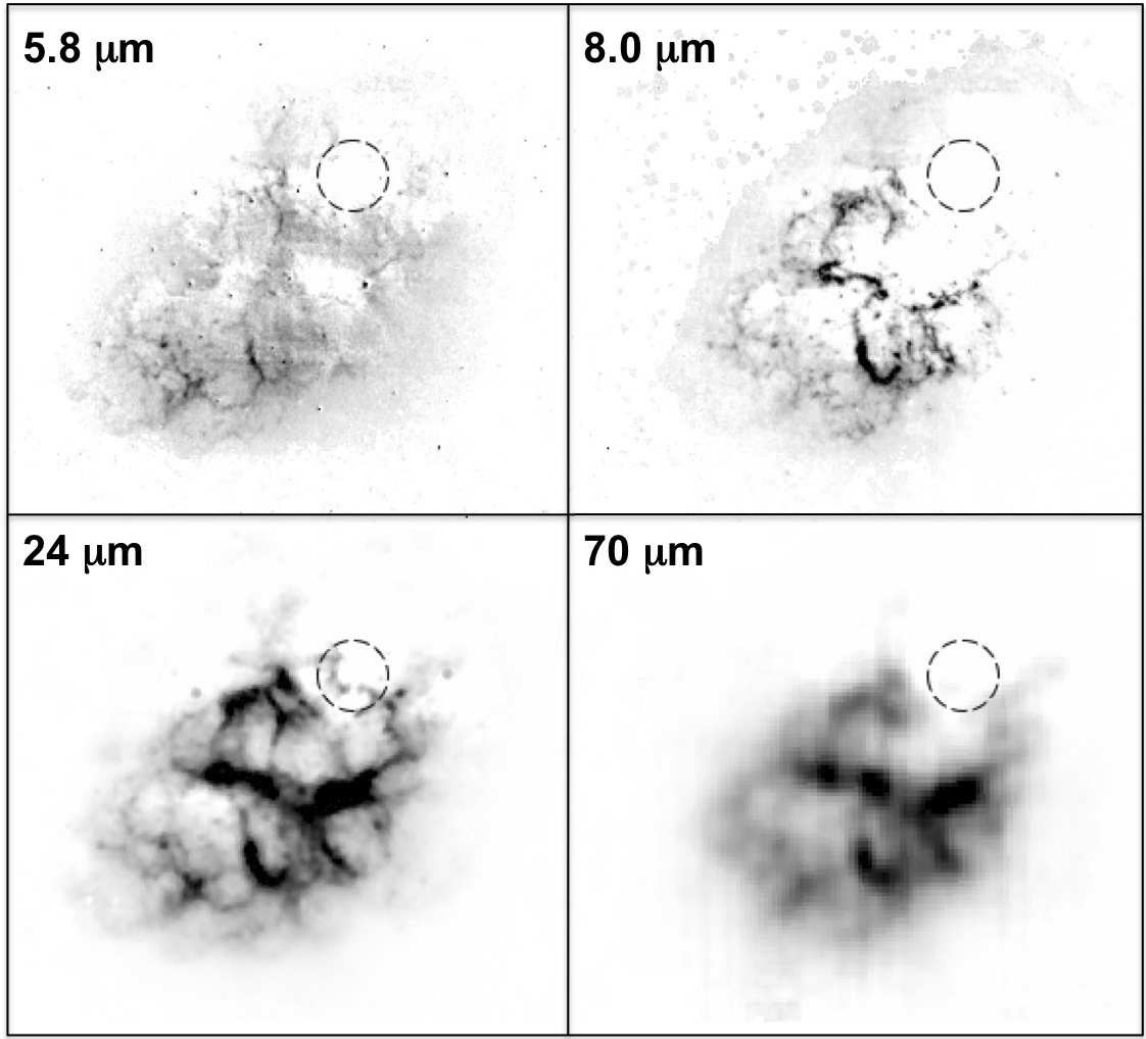} \caption{\label{subsynchmap}Synchrotron-subtracted images of the Crab Nebula. The IRAC 5.8 and 8.0 \micron\ images are on a linear gray scale from 0.0--10 MJy/sr, and the MIPS 24 and 70 \micron\ images are on a linear scale from 0--200 MJy/sr. The dashed circle indicates the region where the extrapolated synchrotron emission is overestimated and where the residuals are negative.}
\end{figure*}

Here, the subscript ``i'' refers to the wavelength of the image from which the extrapolated synchrotron emission is being subtracted. In each case, the 3.6 \micron\ image was convolved to the resolution of $\lambda_i$ \citep{gor08}, and each of the images was extinction corrected using the extinction curve of \citet{chi06}. The IRAC surface brightness correction factors were applied to all IRAC images (Table 4.9 of the IRAC Instrument Handbook). The resulting synchrotron-subtracted images are shown in Figure~\ref{subsynchmap} and will be discussed in Section \ref{morphology}.

Next, we subtracted the synchrotron contribution from the SL and LL spectra for each spatial position. We created four synchrotron spectral cubes covering the wavelength range of each of the four low resolution IRS modules for every spatial pixel along the slits shown in Figure~\ref{slits}. We first checked the calibration of the IRS spectra by comparing the LL emission integrated over the MIPS 24 \micron\ bandpass to the flux measured in the MIPS 24 \micron\ image. We found the normalization factor between the LL spectra and the 24 \micron\ surface brightness to be 0.997. There was no need for an additional scaling factor for the SL spectra, since the continuum levels in the overlapping wavelength regions of the LL and SL spectra matched well. We also integrated this final spectrum over the IRAC 8.0 \micron\ bandpass and found that the flux does not differ significantly ($<$ 5\%) from the flux in the IRAC 8 \micron\ image. We mapped the 3.6 \micron\ image to match the spectral cube map for each of SL and LL orders, and used equation \ref{eq2} to compute a synchrotron emission cube for each module. These synchrotron cubes were then subtracted from the background-subtracted IRS spectral cubes that have been extinction corrected using the extinction curve of \citet{chi06}. This resulted in SL and LL synchrotron-subtracted spectral cubes that only include line emission and continuum emission from dust.

\subsection{Residual Dust Emission}

\begin{deluxetable*}{lcccc}
\tablecolumns{5} \tablewidth{0pc} \tablecaption{\label{linefits}IRS LINE FITS}
\tablehead{
\colhead{Line ID} & \colhead{Wavelength} & \colhead{Region 1 Flux} & \colhead{Region 2 Flux} &  \colhead{Region 3 Flux} \\
\colhead{} & \colhead{(\micron)} & \colhead{($10^{-11} W/cm^{2}/sr$)}  & \colhead{($10^{-11} W/cm^{2}/sr$)}  & \colhead{($10^{-10} W/cm^{2}/sr$)} 
}
\startdata

$[$\ion{Fe}{2}$]$ & 5.340 & 3.897  $\pm$ 0.299 &  1.937 $\pm$ 0.239 & 1.579 $\pm$ 0.174 \\
$[$\ion{Mg}{5}$]$ & 5.610 & \nodata & 0.229 $\pm$ 0.040 & \nodata \\
$[$\ion{Ni}{2}$]$  & 6.636 & 2.474  $\pm$ 0.170 & 0.948 $\pm$ 0.099 & 0.712 $\pm$ 0.062 \\
$[$\ion{Ar}{2}$]$ & 6.985 & 16.72 $\pm$ 0.457 & 9.753 $\pm$ 0.180 & 5.689 $\pm$ 0.150 \\
$[$\ion{Na}{3}$]$ & 7.318 & 0.365 $\pm$ 0.086 & \nodata & \nodata \\
$[$\ion{Ar}{3}$]$ & 8.991 & 3.308 $\pm$ 0.007 & 2.093 $\pm$ 0.526 & 1.481 $\pm$ 0.039 \\
$[$\ion{S}{4}$]$ & 10.511 & 2.372 $\pm$ 0.204 & 2.256 $\pm$ 0.100 & 1.593 $\pm$ 0.055 \\
$[$\ion{Ni}{2}$]$ & 10.682 & 0.222 $\pm$ 0.004 & 0.010 $\pm$ 0.006 & \nodata \\
$[$\ion{Ni}{2}$]$ & 11.308 & 0.067 $\pm$ 0.007 & \nodata & \nodata \\
$[$\ion{Ne}{2}$]$ & 12.814 & 9.383 $\pm$ 0.350 & 46.05 $\pm$ 1.14 & 19.32 $\pm$ 0.41 \\
$[$\ion{Ne}{5}$]$/$[$\ion{Cl}{2}$]$ & 14.322/14.368 & 0.956 $\pm$ 0.220 & 4.320 $\pm$ 1.501 & 1.831 $\pm$ 0.299 \\
$[$\ion{Ne}{3}$]$ & 15.555 & 7.565 $\pm$ 0.123 & 34.32 $\pm$ 0.038 & 18.77 $\pm$ 0.318 \\
$[$\ion{Fe}{2}$]$ & 17.936 & 0.560  $\pm$ 0.010 & 0.491 $\pm$ 0.011 & 0.441 $\pm$ 0.027 \\
$[$\ion{S}{3}$]$ & 18.713 & 4.278  $\pm$ 0.042 & 4.064 $\pm$ 0.137 & 4.895 $\pm$ 0.308 \\
$[$\ion{Fe}{3}$]$ & 22.925 & 0.322  $\pm$ 0.024 & 0.524 $\pm$ 0.046 & 0.268 $\pm$ 0.037 \\
$[$\ion{Ne}{5}$]$/$[$\ion{Fe}{2}$]$ & 24.318/24.519 & 0.281 $\pm$ 0.034 & 1.215 $\pm$ 0.087 & 0.924 $\pm$ 0.047 \\
$[$\ion{O}{4}$]$/$[$\ion{Fe}{2}$]$ & 25.890/25.988 & 3.913 $\pm$ 0.010 & 11.08 $\pm$ 0.34 & 8.739 $\pm$ 0.278 \\
$[$\ion{S}{3}$]$ & 33.481 & 3.876  $\pm$ 0.112 &  3.321 $\pm$ 0.097 & 4.194 $\pm$ 0.126 \\
$[$\ion{Si}{2}$]$ & 34.815 & 2.684  $\pm$ 0.151 & 2.400 $\pm$ 0.125 & 3.057 $\pm$ 0.143 \\
$[$\ion{Fe}{2}$]$ & 35.349 & 0.243  $\pm$ 0.035 & \nodata & \nodata \\
$[$\ion{Ne}{3}$]$ & 36.014 & 0.510  $\pm$ 0.014 & 2.397 $\pm$ 0.134 & 1.279 $\pm$ 0.117 \\
\enddata
\tablecomments{The listed uncertainties are 1-$\sigma$ statistical uncertainties from the fit only.}

\end{deluxetable*}

Inspection of the synchrotron-subtracted spectral cube reveals residual continuum emission from dust in some regions of the nebula. In order to determine the spatial distribution of the dust emission in the IRS slits, we used the synchrotron-subtracted cube to produce a map of the rising continuum. We selected a line-free region of the spectrum and produced a map of the integrated emission between  27--32 \micron. This map is shown in Figure~\ref{dustmap} in magenta, overlaid on the MIPS 24 \micron\ image in blue. The map shows that the brightest dust emission coincides with the brightest filaments in the 24 \micron\ image. We note that the background continuum in the vicinity of the Crab Nebula falls off longward of 25 \micron, so the background contamination in the excess dust continuum is minimal.

To extract spectra for analysis of the dust emission, we chose three positions on the dust map that provided the best signal-to-noise. These regions are shown in Figure~\ref{dustmap}, and also correspond to positions 2, 4, and 5 in Table \ref{obs}. While the LL spectra were extracted from regions outlined in black, the SL spectra were extracted from smaller sub-regions that overlap with the LL slits (see Figure~\ref{slits}). In order to fit the dust emission only, we extracted spectra from the synchrotron-subtracted spectral cubes and measured the MIPS 70 \micron\ flux density for these same regions. We fit the data with a modified blackbody model with three different grain compositions. The best-fit dust models and the unsubtracted spectra are shown in Figure~\ref{dustspec}. The details of the fitting are discussed in Section \ref{dustfitting}.

\subsection{Line Emission}

Strong forbidden emission lines are present in the IRS spectrum at all six positions in the Crab Nebula. We measured line intensities in the sample spectra from three positions in Figures \ref{dustmap} and \ref{dustspec}. The measured line intensities are listed in Table \ref{linefits}. The strongest line emission comes from sulfur, neon, argon, oxygen and iron. Other relatively bright lines include silicon and nickel. There are some obvious differences in the relative line intensities between Region~1, the brightest filament in the south, and Regions 2 and 3, the brightest filaments in E/W direction. While the southern filament shows stronger emission from [\ion{Ar}{2}], [\ion{Ar}{3}], [\ion{Na}{3}], and [\ion{Ni}{2}], the east and west filaments show stronger emission from [\ion{Ne}{2}], [\ion{Ne}{3}] and [\ion{O}{4}]. All three regions show similarly bright emission from [\ion{S}{3}], [\ion{S}{4}], [\ion{Si}{2}], and [\ion{Fe}{2}]. 
Emission from lines with higher ionization potentials, such as [\ion{O}{4}], [\ion{Ne}{5}], and [\ion{Mg}{5}], is strongest in Region~2. Comparisons of line intensity ratios for lines of similar ionization potentials reveal some differences across the three regions that may be indicative of variations in the gas abundances. While [\ion{S}{3}] is equally bright in all three regions, [\ion{Ne}{2}] is much more pronounced in Region~2. The [\ion{Ni}{2}]/[\ion{Si}{2}] line ratio is more than a factor of two higher in Region~1, and [\ion{Na}{3}] is not even detected in Region~2, where [\ion{Ne}{3}] and [\ion{O}{4}] are most prominent. 

Since the velocity splitting of the line emission from the two sides of the expanding nebula is not resolved in SL and LL modules, we cannot make any definitive conclusions about the variations in gas properties based on the low-resolution data. A detailed analysis and modeling of the high-resolution line emission detected by \textit{Spitzer} will be presented in a subsequent publication. In this paper, we use the measured line intensities to determine the contribution of line emission to the total integrated flux from the Crab Nebula in order to estimate the total mass of dust.

\section{Morphology of the Dust and Line Emission} \label{morphology}

\begin{figure}
\epsscale{1.15} \plotone{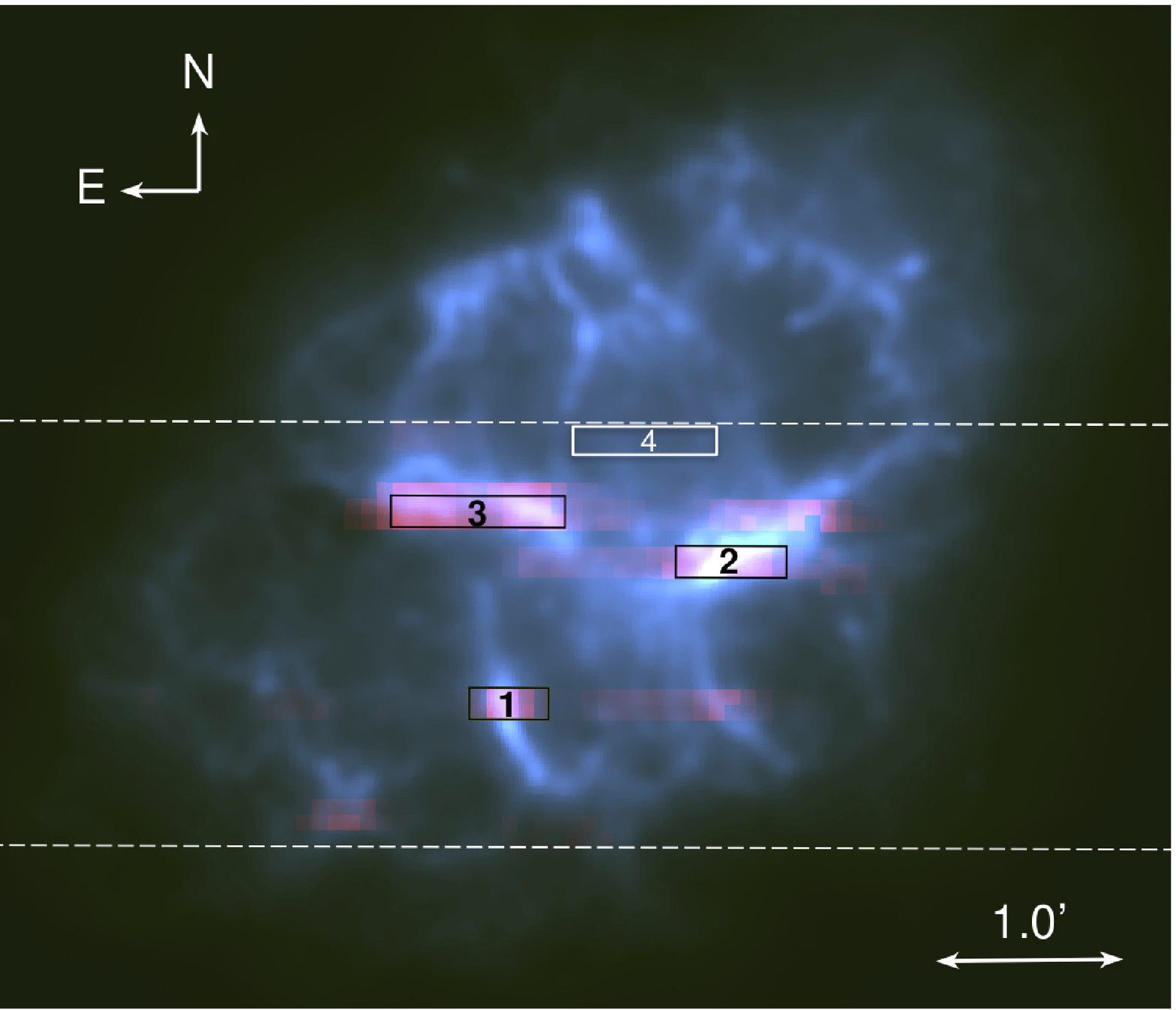} \caption{\label{dustmap}Map of the synchrotron-subtracted line-free region between 27--32 \micron, representing continuum emission from dust in the LL slits (magenta) overlaid on the MIPS 24 \micron\ image (blue). The total coverage of the LL slits is shown in Figure~\ref{slits}. The southern and northern edges of the LL slits are marked by white dashed lines. The brightest dust emission is seen to spatially coincide with the brightest filaments.The extraction regions for the dust spectra in Figure~\ref{dustspec} are overlaid in black and numbered.}
\end{figure}

\begin{figure} 
\epsscale{1.18} \plotone{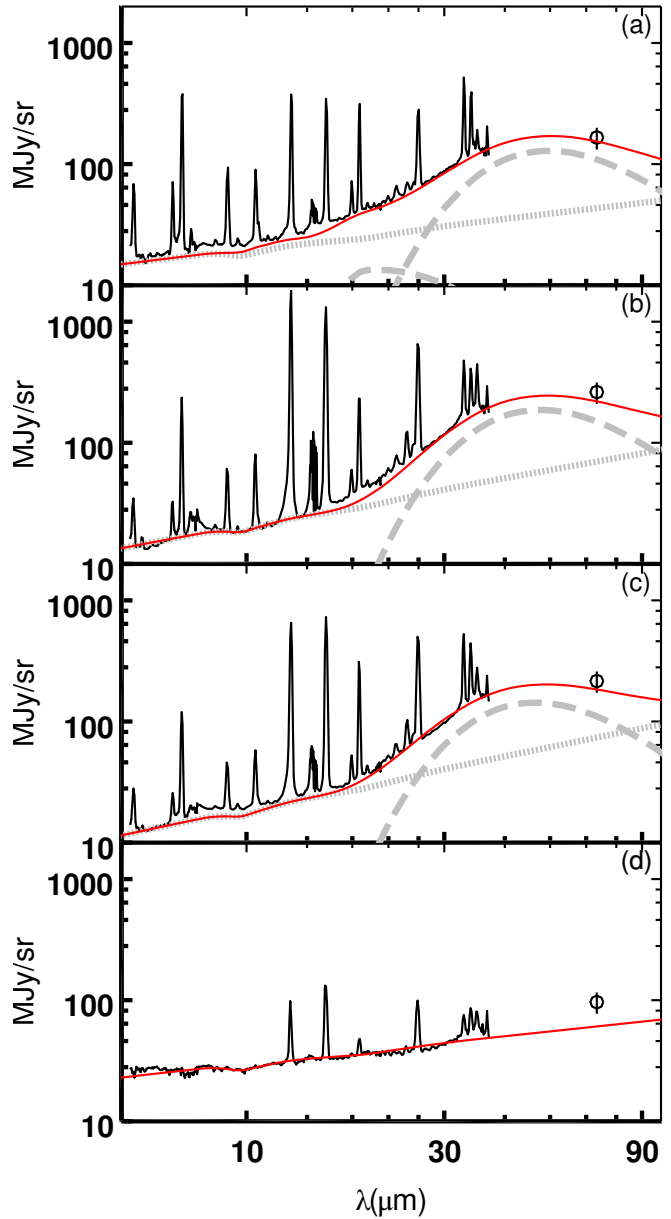} \caption{\label{dustspec}Model fits to the spectral continuum and the 70 \micron\ data point for three extraction regions shown in Figure~\ref{dustmap}. Spectra from regions 1, 2, 3, and 4 are shown in plots (a), (b), (c), and (d), respectively. The model is silicate dust (dashed line), plus a synchrotron component whose index and normalization have been fixed to the values derived from the IRAC data (dotted line). The red line represents the best fit two-component model corresponding to a best-fit grain temperature of 53 K, 55 K, and 57 K, for plots (a), (b), and (c). There is an additional dust component with a temperature of $\sim$ 136 K in Region~1. The amorphous carbon grains provide equally good fits to the data. The best-fit values for both compositions are listed in Table \ref{dustfitstab}. For comparison, the spectrum in plot (d) is from a region centered on the PWN that does not show a significant excess from dust. The red line in plot (d) is the synchrotron component only, extrapolated from the 3.6 \micron\ image.}
\end{figure}

The synchrotron-subtracted images of the Crab Nebula are shown in Figure~\ref{subsynchmap}, while the unsubtracted images are shown in Figure~3 of \citet{tem06}. The synchrotron component that was extrapolated from the IRAC 3.6 \micron\ image describes the synchrotron emission at the longer IRAC and MIPS wavelengths fairly well. The synchrotron emission from the torus and jet features is completely subtracted out by our method. The extrapolation overestimates the synchrotron contribution in the NW part of the nebula, in a region shown by the dashed circle in Figure~\ref{subsynchmap}. The residual IRAC images show slightly negative values in the NW, and this overestimate increases with wavelength, with the most significant over-subtraction at 70 \micron. The over-subtracted values in the NW are on the order of $\sim$ 1 MJy/sr at 5.8 \micron, 3 MJy/sr at 8.0 \micron, 5--10 MJy/sr at 24 \micron, and 15--25 MJy/sr at 70 \micron. These values range from approximately 5--20 \% of the total extrapolated synchrotron flux in this region, and do not appear to be caused by the uncertainties involved in the subtraction. 

The uncertainty in the spectral index is dominated by the calibration uncertainties in the IRAC flux density and extinction correction at 3.6 and 4.5 \micron, so it is expected to be a systematic uncertainty with no significant spatial variations. The over-subtraction from the extrapolated synchrotron spectrum is seen in one localized region of the nebula, and this suggests that it may be caused by an intrinsic difference in the synchrotron spectrum in this region. As the magnetic field evolves in the expending nebula, we might expect a curvature in the PWN's power-law spectrum. A slight curvature of the spectrum in the IR may explain why the synchrotron over-subtraction increases with wavelength, but we still do not understand why the curvature may be different or more pronounced in NW part of the nebula in particular.

The residual filamentary emission is dominated by the lines of [\ion{Fe}{2}] (5.34 \micron) in the IRAC 5.8 \micron\ image, [\ion{Ar}{2}] (6.99 \micron) in the IRAC 8.0 \micron\ image, [\ion{O}{4}] (25.89 \micron),  [\ion{Fe}{2}] (25.99 \micron) and dust emission in the MIPS 24 \micron\ residual image, and mostly dust emission in the MIPS 70 \micron\ image with some contribution from [\ion{O}{1}] (63.18 \micron) and [\ion{O}{3}] (88.36 \micron) lines. We integrated the IRS spectra extracted from the synchrotron-subtracted spectral cube at the three positions in Figure~\ref{dustmap} over the MIPS 24 \micron\ bandpass. We repeated this integration for line-subtracted spectra in order to compare the values and estimate the contribution of line emission in the MIPS 24 \micron\ image. We find that approximately 27\% of the residual emission comes from lines in Region~1, 54\% in Region~2, and 48\% in Region~3. This suggests that on average, about half of the emission in the MIPS 24 \micron\ synchrotron-subtracted image comes from dust.

The IRS spectral map of the dust continuum overlaid on the MIPS 24 \micron\ image is shown in Figure~\ref{dustmap}. The dust emission spatially correlates with the ejecta filaments, with the brightest dust emission concentrated along the brightest filaments. This is consistent with previous optical studies of the Crab Nebula that found a correlation of dust absorption features with the filament cores \citep{fes90,bla97,hes90}.

\begin{deluxetable*}{lccccc}
\tablecolumns{6} \tablewidth{0pc} \tablecaption{\label{dustfitstab}Dust Temperature and Mass}
\tablehead{
  \colhead{Composition} &  \multicolumn{3}{c}{Best-Fit Temperature (K)} &  \colhead{Average Dust} & \colhead{Total Dust Mass in} \\
  \colhead{} & \colhead{Region 1} & \colhead{Region 2} & \colhead{Region 3} & \colhead{Temperature (K)} & \colhead{ the Crab Nebula ($\rm M_{\odot}$)}
  }
\startdata
Silicates & 53 $\pm$ 3 (136 $\pm$ 12) & 55 $\pm$ 2 & 57 $\pm$ 2 & 55 $\pm$4  & $(2.4^{+3.2}_{-1.2})\times10^{-3}$ \\
Carbon & 57 $\pm$ 6 (119 $\pm$ 12) & 61 $\pm$ 3 & 63 $\pm$ 2 & 60 $\pm$ 7 & $(3.2^{+8.7}_{-2.1})\times10^{-3}$ 
\enddata
\tablecomments{The temperature uncertainties in the three regions represent 1.6 $\sigma$ statistical uncertainties from the fits. The total mass values have been calculated by assuming that all of the dust in the Crab Nebula emitting at 24 \micron\ is at the average observed dust temperature. In Region~1, the temperatures in parentheses are best-fit temperatures for the hot dust component that comprises less than 1\% of the total mass in that region.}
\end{deluxetable*}

\section{Dust Composition and Temperature}\label{dustfitting}

The extracted IRS spectra are shown in Figure~\ref{dustspec} and they clearly show a rising dust continuum above the synchrotron power law. The residual dust emission shows no evidence for PAH features or for any obvious silicate emission features, other than those caused by the extinction along the line of sight. The dust spectrum is featureless and shows no emission features typical of ejecta dust in other young remnants \citep{rho09}.

In order to model the residual dust emission, we first corrected the spectrum for extinction along the line of sight using the mid-IR extinction curve from \citet{chi06}. The extinction can be observed as the dip in the synchrotron power-law spectrum around 10 \micron. We then fitted the synchrotron-subtracted spectra with a dust emission model using two different grain compositions; amorphous carbon \citep{zub04} and silicates \citep{wei01}. Both compositions produced equally good fits to the observed spectra and the 70 \micron\ data point. The silicate composition requires an average temperature of 55 $\pm$ 4 K, and amorphous carbon grains an average temperature of 60 $\pm$ 7 K.  We do not see significant changes in temperature between the three regions. 

There appears to be some excess emission between 20--30 \micron\ in regions 1 and 2 that suggests that an additional warmer dust component with a much smaller mass may be contributing to the spectrum. While the addition of this second component did not significantly improve the fit for Region~2, an additional component with a temperature of 136 $\pm$ 12 K for silicates, and 119 $\pm$ 12 K for carbon grains did improve the fit in Region~1. If present, this warmer component represents only a tiny fraction of the total dust mass, on the order of 0.2 \%. Such emission may be explained by stochastically heated smaller grains. Since the spectrum below 15 \micron\ is dominated by synchrotron emission, the SL data were not sensitive enough to constrain the emission from PAHs and the small, stochastically heated grains.

The best fit dust temperatures for each region and each grain composition are listed in Table \ref{dustfitstab}. The best fit dust models are shown in Figure~\ref{dustspec}. As described in Section \ref{synchanalysis}, the power-law component that is added to the models in Figure~\ref{dustspec} was derived from the synchrotron index map in Figure~\ref{indexmap} and the synchrotron-dominated IRAC 3.6 \micron\ image. The spectral index $\alpha$ (normalization) in the derived power-law component shown in Figure~\ref{dustspec} is 0.4 (15.0 MJy/sr) in Region~1, 0.6 (11.8 MJy/sr) in Region~2, 0.7 (9.7 MJy/sr) in Region~3, and 0.36 (21.9 MJy/sr) in Region~4. The normalization value is the surface brightness of the synchrotron component at 3.6 $\micron$ measured in each extraction region.

\section{Dust Heating}

\begin{figure*} 
\epsscale{1.15} \plottwo{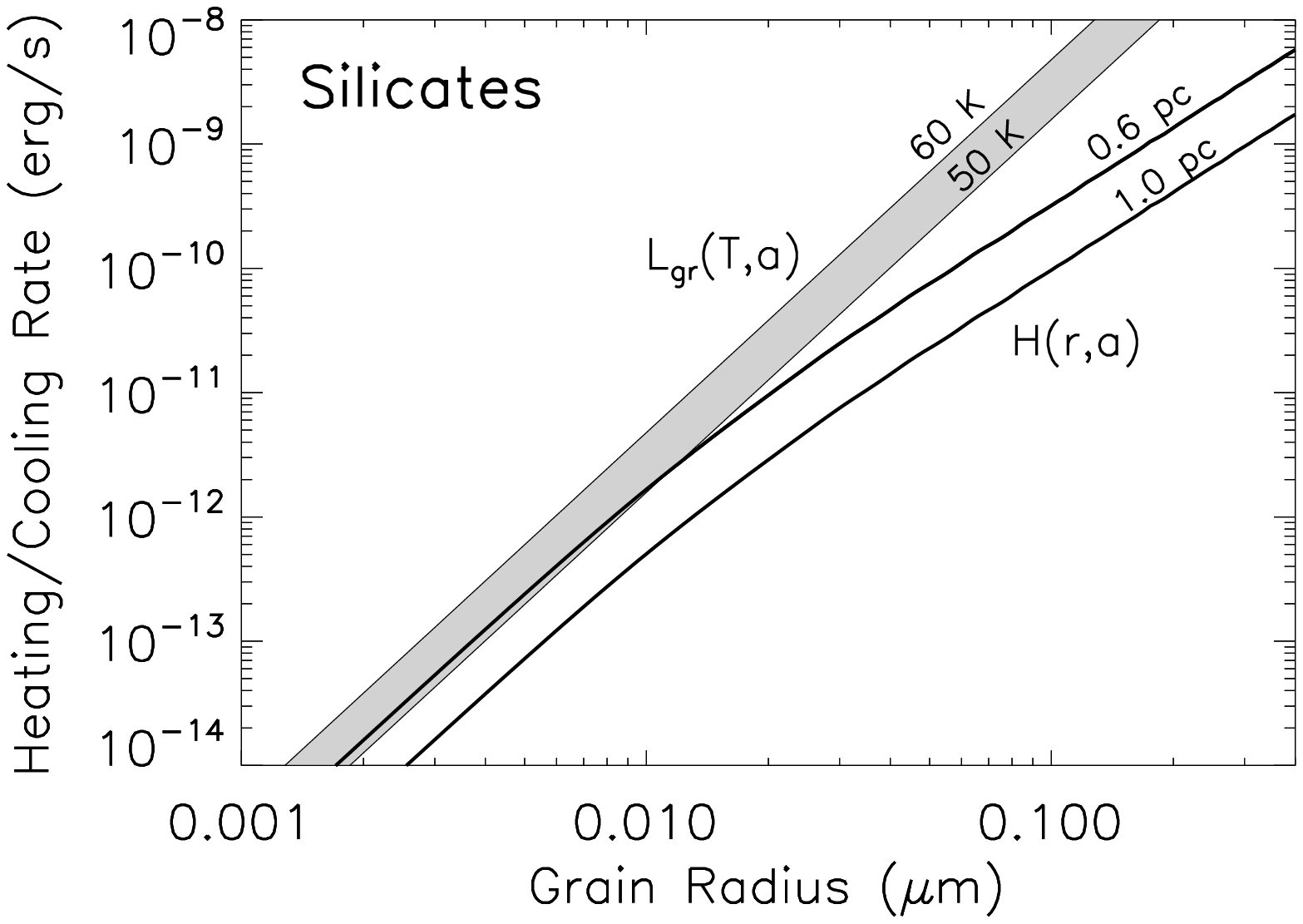}{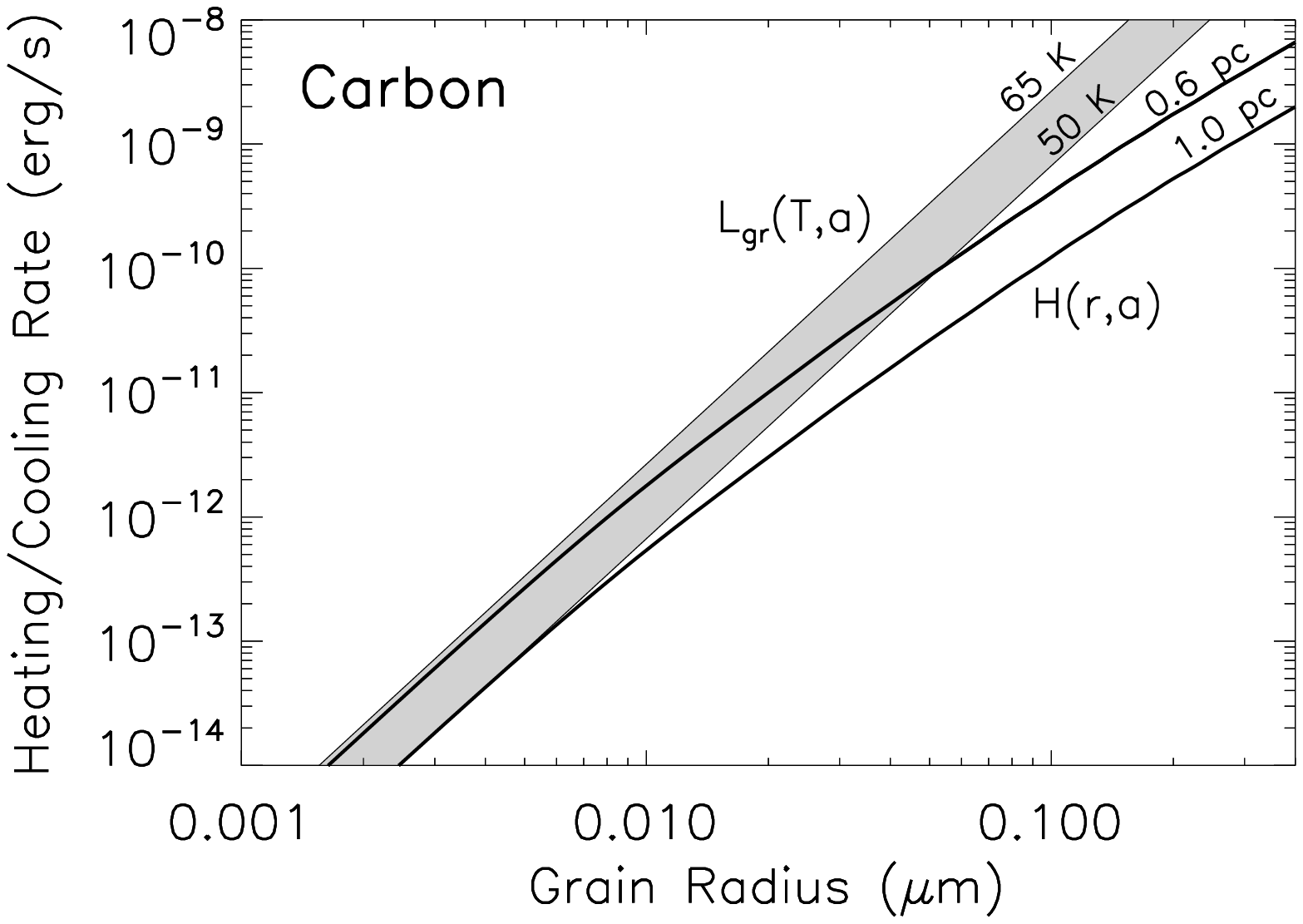} \caption{\label{heatrate} The heating rate of dust by the PWN's broadband synchrotron spectrum, as a function of grain size, for a source distance ranging from 0.6--1.0 pc \citep{cad04} is shown by the solid curves. The cooling rate of dust at the observed, best-fit temperature is shown by the gray band, where the width of the band represents the uncertainties in the temperature measurements listed in Table \ref{dustfitstab}. The place where the heating and cooling rates cross gives the expected grain radius of the emitting dust.}
\end{figure*}

The dominant heating source for the dust in the Crab Nebula is synchrotron radiation from the pulsar wind, while collisional heating by the gas in the filaments plays a smaller role \citep{dwe81}. In order to determine if the PWN can heat the dust to the observed temperature, we calculated the expected dust heating rate by the PWN in the Crab Nebula and compared it to the cooling rate of dust at the observed temperatures. We assume that the filaments are optically thin to the incoming radiation. A rough estimate of the optical depth gives $\tau \leq 1$, assuming a filament size of $\sim 10^{16} \: \rm cm$, a density of 2000 $\rm cm^{-3}$ and a dust-to-gas mass ratio of 0.1 \citep{san98}. Because of the low $\tau$, we neglected any internal absorption in the nebula. The radiation absorbed by a dust grain is then given by the heating rate
\begin{equation}
H=\frac{4\pi a^2 \int L_{\nu}Q(\nu,a)d\nu}{4\pi r^2},
\end{equation}
where $L_{\nu}$ is the non-thermal specific luminosity of the Crab Nebula's PWN at a given frequency, $a$ is the grain size, $Q(\nu,a)$ is the frequency and grain size dependent absorption coefficient for a given grain composition, and $r$ is the distance between the radiation source, i.e. the pulsar, and the dust grain, assumed to be between 57\arcsec--100\arcsec, or 0.55--1.0 pc for a distance of 2 kpc. This distance range was chosen based on the location of the ejecta filaments in 3-dimensional models of the Crab Nebula \citep{cad04}. The non-thermal broadband spectrum of the Crab Nebula summarized in \citet{hes08} was used for the heating luminosity $L_{\nu}$. In calculating the heating rate of the dust, we integrated the product of $L_{\nu}$ and the dust absorption efficiencies up to an energy of about 0.6~keV. The fraction of the energy deposited in the dust at higher energies makes a negligible contribution of less than 1~\% to the heating of the dust because of the combined effects of decreasing $L_{\nu}$ with energy, and the decreasing efficiency of the energy deposited by the photoelectrons in the dust \citep{dwe96}.

The radiative cooling rate for a dust grain is given by 
\begin{equation}
L_{gr}=4\pi a^2\int\pi B_{\nu}(T) Q(\nu,a)d\nu,
\end{equation} 
where $B_{\nu}$ is the Planck function and $T$ is the dust grain temperature \citep{dwe81, dwe08}. 
In equilibrium, the heating rate $H(r,a)$ and the cooling rate $L_{gr}(T,a)$ should balance each other. Since the distance from the heating source $r$ and the dust temperatures $T$ are both measured parameters, and since both rates also depend on the grain radius $a$, we can calculate a grain size for which $H(r,a)=L_{gr}(T,a)$. This quantity then represents the grain size that is required in order for the PWN to heat the dust to the observed dust temperatures in Table \ref{dustfitstab}. 

Due to the uncertainties in the fitted dust temperatures and the spread in the possible distance to the heating source that ranges from 0.55--1.0 pc, there is actually a range of possible dust grain radii for each grain composition. This is demonstrated in Figure~\ref{heatrate} that shows the heating and cooling rates as a function of grain radius for silicate and amorphous carbon grains. The solid black lines represent the heating rate at the two distance extremes, while the gray band represents the cooling rate for the spread of best-fit dust temperatures. The overlapping regions of the heating and cooling bands represent the possible dust grain sizes.

It can be seen from Figure~\ref{heatrate} that the silicate dust would need to be located at the closest distance extreme from the heating source in order to be heated to even the low-end of the observed temperatures. Since the filaments are distributed at some range of distances from the center of the Crab Nebula's PWN, the grain composition may mostly be carbonaceous, or at least a mixture of silicate and carbonaceous material. Theoretical models based on classical nucleation and grain growth theory suggest that carbon dust is the most abundant dust species in Type IIP SNe with lower mass progenitors \citep{koz09}, so a carbonaceous composition would not be unusual for the Crab Nebula, which is thought to have been produced in a Type IIP explosion with a 8--12 $M_{\odot}$ progenitor \citep[e.g.,][]{dav85,che05,mac08}.

The most interesting thing to notice in Figure~\ref{heatrate} is that the required grain radii are fairly small for both silicate and carbon grain compositions. The required grain radii are $< 0.015\: \micron$ for silicate grains, and  $< 0.05\: \micron$ for amorphous carbon grains. Dust formation takes place in the rapidly expanding SN ejecta that are internally heated by radioactivties.The composition and sizes of the grains depend in a complex way on the kinematics, thermodynamic history, and composition of the ejecta, as well as the initial number of nucleation centers around which the dust grows. The studies of dust properties in SNRs can therefore be used to relate the inferred grain sizes to the complex physical processes in the ejecta and provide some insight into the properties of the progenitor star.

\citet{koz09} and \citet{noz10} have investigated the dependence of dust formation in SN ejecta on the type of core collapse SN. They compared the mass and properties of dust formed in Type IIP SNe with massive H-envelopes and Type IIb SNe with low mass H-envelopes and found that the total dust mass of 0.1-0.7 $\rm M_{\odot}$ does not depend on the thickness of this outer envelope, but that the grain size is strongly dependent on its thickness. A less massive envelope allows for a much higher expansion velocity of the He-core, leading to a much lower ejecta gas density, too low for the larger grains to form. For example, \citet{koz09} find that the radii of dust grains for a 0.08 $\rm M_{\odot}$ H-envelope are below 0.006 \micron, while the radii of grains formed in a SN with a 10 $\rm M_{\odot}$ H-envelope are larger than 0.03 \micron.  
The low initial mass of the Crab Nebula's progenitor and the relatively low expansion velocity of the optical filaments suggest that the Crab Nebula is a result of a Type IIP SN \citep[e.g.][]{che05,mac08}. The models would therefore predict somewhat larger grain radii then inferred from our observations, unless the Crab's progenitor experienced substantial mass loss prior to the SN event. 

A low dust grain size also has important implications for the survival of SN dust and its injection into the ISM. The majority of dust grains with small radii are expected to be destroyed by sputtering in the SN reverse shock \citep{dwe05,koz09,bia09,noz10}. Hydrodynamic simulations of grain destruction by SNR reverse shocks show that grains with initial radii below 0.1 \micron\ are almost completely destroyed \citep{sil12}. This implies that the majority of the dust observed in the Crab Nebula by \textit{Spitzer} should eventually be destroyed in the reverse shock. However, it is possible that clumping of gas and dust in the Crab's filaments may provide some shielding for the dust grains.

\section{Dust Mass Estimate}

In order to estimate the total mass of dust in the Crab Nebula, we first estimated the contribution of dust emission to the total integrated flux. After subtracting the synchrotron emission extrapolated from the IRAC 3.6 \micron\ image (Figure~\ref{subsynchmap}), the total residual contribution from dust and line emission at 24 \micron\ is 24.3 $\pm$ 2.4 Jy, or about 40\% of total flux. As described in Section \ref{morphology}, we calculated the fraction of line emission in three regions shown in Figure~\ref{dustmap}. We find that the dust continuum emission accounts for 73\%, 46\%, and 52\% of the total residual emission in regions 1, 2, and 3, respectively. In order to estimate the dust fraction over the entire nebula, we assume an average value of 57\% of dust emission, which gives 13.9 Jy for the total dust emission in the nebula at 24 \micron. We use this value, and the average of the best-fit temperatures for the three regions (Table \ref{dustfitstab}) to approximate the total mass of dust in the Crab Nebula. We used the equation
\begin{equation}
{M}_{dust}=\frac{{F}_{\nu}(\nu)d^{2}}{{B}_{\nu}(\nu,T)\kappa_{\nu}(\nu)},
\end{equation}
where $F_{\nu}(\nu)$ is the total IR flux at 24 \micron, $\kappa_{\nu}(\nu)$ is the grain opacity equal to $\frac{3Q(\nu,a)}{4{\rho}a}$, $d$ is the distance, $B_{\nu}(\nu,T)$ is the Planck function evaluated at the grain temperature, and $\rho$ is the grain density. We used the opacity values for silicate and amorphous carbon grains summarized in \citet{zub04}. The dust mass values for both grain compositions are listed in Table \ref{dustfitstab}. The total estimated mass for silicate grains is $(2.4^{+3.2}_{-1.2})\times10^{-3}\rm \:M_{\odot}$, and the total mass for amorphous carbon grains is $(3.2^{+8.7}_{-2.1})\times10^{-3}\rm \:M_{\odot}$. We note that these estimates assume a single average dust temperature over the entire nebula, and that the actual composition of the dust is likely to be a mixture of the two composition. We estimate the total dust mass in the Crab Nebula to be $(1.2-12)\times10^{-3}\rm \:M_{\odot}$ depending on the assumed composition.

The derived mass values are well below the theoretical predictions for Type II SNe. Classical nucleation theory \citep{noz03,koz09} and the theory of chemical kinematic approach for the formation of molecular precursors \citep{che10} predict that each SN should form 0.1--0.7 $\rm M_{\odot}$ of dust. Type IIP SNe should form on the order of 0.1 $\rm M_{\odot}$ of dust, at least an order of magnitude more than what we observe in the Crab Nebula. Furthermore, the small grain radii inferred from the observations indicate that much of the observed dust will eventually be destroyed by the SN reverse shock. It is possible that dust grains with a colder temperature reside outside of the visible Crab Nebula's radius. These grains would emit mostly at far-IR and sub-mm wavelengths, and observations with the \textit{Herschel Space Observatory} and the Atacama Large Millimeter/sub-millimeter Array (ALMA) should provide a constraint on the possible colder dust mass component in the Crab nebula.

\section{CONCLUSIONS}
 
In this paper, we have analyzed the \textit{Spitzer Space Telescope} low-resolution spectroscopy of several regions across the Crab Nebula with the goal of isolating the emission from SN produced dust and determining its physical properties and spatial distribution. The dust emission is spatially coincident with the brightest filaments in the ejecta. The dust spectrum appears featureless, with no obvious PAH emission features, or other dust emission features typical of ejecta dust recently observed in other SNRs. We fitted the dust emission spectra with dust emission models of two different grain compositions; silicates and amorphous carbon grains. Both grain compositions provide equally good fits to the observed emission. The best-fit model yields an average temperatures of 55 $\pm$ 4 K for silicates, and 60 $\pm$ 7 K for amorphous carbon grains. An emission excess between 20--30 \micron\ suggest that a very small amount of warmer dust is present in some regions, possibly as a result of stochastic heating of smaller grains. 

The dust in the Crab Nebula is predominately heated by the surrounding synchrotron radiation field, rather than collisionally heated by the gas. Our heating models show that the dust grain radii need to be relatively small in order for the dust to be heated to the observed temperature. The grain radii in the Crab Nebula would need to be $<$ 0.015 \micron\ for silicates and $<$ 0.05 \micron\ for carbon grains, unlike what is expected in Type IIP SNe. Assuming that the dust is composed predominantly of silicates or carbon, the total dust mass produced in the ejecta of the Crab is estimated to be $(1.2-12)\times10^{-3}\:M_{\odot}$, at least an order of magnitude less than was is expected from theoretical predictions. Due to the small grain size, much of the dust may eventually be destroyed by sputtering in the SN reverse shock.

Existing \textit{Herschel} observations and future observations with ALMA will provide constraints on the mass of colder and larger dust grains that may be present in the Crab Nebula. The orders of magnitude better sensitivity and resolution of the \textit{James Webb Space Telescope} (JWST) will allow us to probe the smaller dust grains and PAHs in the Crab Nebula's filaments, which will provide further insight into dust grain destruction and processing.

\acknowledgments
This work is partly based on observations made with the \textit{Spitzer Space Telescope}, which is operated by the Jet Propulsion Laboratory, California Institute of Technology under a contract with NASA. RDG was supported by NASA and the US Air Force.

\end{document}